\documentclass[prd,preprint,tightenlines,floatfix,preprintnumbers,nofootinbib]{revtex4}
\usepackage[dvips,final]{graphicx}
\begin{document}

\thispagestyle{empty}

\title{Conformal symmetry based   relation between Bjorken  and  
Ellis-Jaffe sum rules}

\author{A.~L.~Kataev}
\email{kataev@ms2.inr.ac.ru}
\affiliation{Institute for Nuclear
Research of the Russian Academy of Sciences,  117312 Moscow, Russia}

\vspace {10mm}
\begin{abstract}
The identity  between perturbative  expressions for the  coefficient 
functions of the   Bjorken and Ellis-Jaffe 
sum rules is derived in the conformal invariant 
limit of  massless $U(1)$ theory, i.e.  in the perturbative 
quenched QED model. It is also satisfied  in the conformal invariant limit 
of the massless  $SU(N_c)$  gauge theory  with fermions. The latter limit  
is realized  in the imaginable case, when all perturbative coefficients 
of the corresponding renormalization group  $\beta$-function are 
equal to zero.    
The derivation is    
based on the  comparison of results of 
application of the operator product  
expansion approach to  the dressed  triangle  
Green functions of singlet Axial vector-  Vector-Vector and non-singlet  
Axial vector-Vector-Vector fermion currents in the limit, when the conformal 
symmetry remains unbroken. 
The expressions for  the $O(\alpha_s^3)$ approximation of  the 
non-singlet coefficient 
function, derived in the conformal 
invariant limit   of $SU(N_c)$ group, 
is  reminded. Its possible 
application  in  the phenomenological  analysis  of the 
experimental data for the Bjorken polarized sum rule is outlined.
\end{abstract}
\maketitle


\section{Introduction}

The definitions of the  massless perturbative expressions for the Bjorken and 
Ellis-Jaffe sum rules of the polarised lepton-nucleon DIS are well-known 
and have the following form
\begin{eqnarray}
\label{Bjp}
Bjp(Q^2)&=&\int_0^1\big(g_1^{lp}(x,Q^2)-g_1^{ln}(x,Q^2)\big)dx=\frac{1}{6}g_A 
C_{NS}(A_s(Q^2))  \\ \label{EJp}
EJ^{lp(n)}(Q^2)&=&C_{NS}(A_s(Q^2))(\pm\frac{1}{12}a_3+\frac{1}{36}a_8)+
C_{SI}(A_s(Q^2))
\frac{1}{9}\Delta\Sigma(Q^2)
\end{eqnarray}
where  $a_3=\Delta u-\Delta d$=$g_A$,  $a_8=\Delta u+\Delta d- 2 \Delta s$,
$\Delta u$, $\Delta d$  and $\Delta s$ are the polarised parton distributions
and the subscript $lp(n)$ indicate the polarised DIS of 
charged leptons ($l$) on protons ($p$) and   
neutrons ($n$).   
In the $SU(N_c)$ colour gauge theory $A_s=\alpha_s/(4\pi)$.
The order $O(A_s^3)$ and $O(A_s^4)$ perturbative 
expressions for the non-singlet (NS) coefficient function 
$C_{NS}(A_s)$ were analytically evaluated  
in   
\cite{Larin:1991tj} and \cite{Baikov:2010je} correspondingly, 
while the analytical expressions 
for the  leading in 
the number of quarks flavours terms (renormalon contributions)  were
obtained in \cite{Broadhurst:1993ru} (see \cite{Broadhurst:2002bi}
as well). The singlet (SI) contribution $C_{SI}$ to Eq.(\ref{EJp})  contains 
the coefficient function, calculated   
in  \cite{Larin:1997qq}   at the 
$O(A_s^3)$- level, while   
the SI  anomalous dimension term is  known analytically  from 
the $O(A_s^2)$ and $O(A_s^3)$ results of 
\cite{Larin:1994dr} and  \cite{Larin:1997qq} respectively.
In all these calculations the $\overline{MS}$-scheme was used. 
In this scheme the polarised gluon distribution 
$\Delta G$ does not enter into  Eq.(\ref{EJp}).  
Our main aim is to prove, that  the analytical 
expressions for $C_{NS}$ and $C_{SI}$, defined in 
Eq.(\ref{EJp}), are identical in all orders of perturbation theory  
in the {\it conformal invariant 
limit}  of the massless  $U(1)$ model 
with fermions, i.e.  in the perturbative quenched QED (pqQED) 
approximation, and in the {\it conformal invariant limit } of the massless 
$SU(N_c)$ gauge model with fermions. The latter limit is   realized in the 
{\it imaginable case}, when all perturbative  coefficients 
$\beta_i$-function  of the renormalization-group $\beta$-function of  
$SU(N_c)$ gauge theory with 
fermions are identically equal to zero.  
           
While proving this identity we follow the pqQED  studies, given in  
\cite{Kataev:2010tm}, where    
the classical  Crewther relation 
\cite{Crewther:1972kn}, derived in the quark-parton era  
from the three-point  Green function of the NS Axial 
vector-Vector-Vector(AVV) currents, is  compared with the similar Crewther-type 
relation, which follows from the  
three-point Green function of singlet Axial vector-Vector-Vector currents.    
In the era of continuing  understanding 
of the special  features of the  relations between NS    
characteristics of strong interactions, evaluated within 
perturbative approach  in the  the $SU(N_c)$ gauge  group 
(see  \cite{Broadhurst:1993ru},
\cite{Baikov:2010je}, \cite{Kataev:2010du}, \cite{Baikov:2012zn}  ),  
the detailed considerations      
of the relations, which follow from  the  three-point  Green functions 
of the NS  
AVV currents,  
were studied theoretically in \cite{Gabadadze:1995ei}, 
\cite{Crewther:1997ux}, \cite{Braun:2003rp}.  
 The comment 
on   possible phenomenological applications  
of the conformal-symmetry motivated expression for the   Bjorken polarised 
sum-rule, which in QCD depends from  the scale,    fixed within {\it  
principle of maximal conformality} \cite{Brodsky:2011ig}, 
\cite{Brodsky:2011ta}, is  given. 

\section{Proof of the identity} 

Theoretical considerations of   \cite{Crewther:1972kn} are based on 
the property that in  the {\it conformal invariant limit} the dressed expression for the  
three-point  Green functions of NS AVV  currents is proportional to the 
{\it 1-loop} expression of the related three-point diagram 
\cite{Schreier:1971um}. In the momentum space this means, that    
\begin{equation}
T_{\mu\alpha\beta}^{abc}(p,q)=i\int<0|TA_{\mu}^{a}(y)
V_{\alpha}^b(x)V_{\beta}^c(0)|0>e^{ipx+iqy}dxdy=
d^{abc}\Delta_{\mu\alpha\beta}^{(1-loop)}(p,q)
\label{anvv}
\end{equation}
where $A_{\mu}^{a}(y)=\overline{\psi}(y)\gamma_{\mu}(\lambda^{a}/2)\gamma_{5}
\psi(y)$ and  
$V_{\alpha}^{b}(x)=\overline{\psi}(x)\gamma_{\mu}(\lambda^{b}/2)\psi(x)$ are  
the NS Axial-vector and Vector  currents.
In the same limit it is possible to write-down the similar expression 
for the three-point Green function of SI  Axial vector-NS Vector-Vector 
currents \cite{Kataev:1996ce}
\begin{equation}
T_{\mu\alpha\beta}^{ab}(p,q)=i\int<0|TA_{\mu}(y)
V_{\alpha}^a(x)V_{\beta}^b(0)|0>e^{ipx+iqy}dxdy=
\delta^{ab}\Delta_{\mu\alpha\beta}^{(1-loop)}(p,q)
\label{asvv}
\end{equation}
where $A_{\mu}(y)=\overline{\psi}(y)\gamma_{\mu}\gamma_{5}\psi(y)$.
Thus, the  cancellation of  one-loop corrections to the 
three-point AVV Green function, which was demonstrated by the explicit 
calculations, preformed in Ref.\cite{Jegerlehner:2005fs} within 
dimensional regularization  \cite{'tHooft:1972fi}, can be understood using the 
concept of the conformal symmetry and  demonstrate the  
validity of the theoretical work of Ref.\cite{Schreier:1971um}.

The 
SI coefficient function of the Ellis-Jaffe sum rule is defined  
as the 
coefficient function of the  SI structure in the operator-product expansion 
of two NS Vector currents,namely  
 \begin{equation}
i\int T V_{\alpha}^{a}(x) V_{\beta}^b(0)e^{ipx} d^4x |_{p^2\rightarrow\infty}\approx 
i\delta^{ab}\epsilon_{\alpha\beta\rho\sigma}\frac{p^{\sigma}}{P^2}
C^{SI}_{EJp}(A_s)~A_{\rho}(0)+ \dots
\label{SI}
\end{equation}
The expression should be compared with the  definition of the
NS coefficient function , which enters into  
operator-product of  the three-point Green function 
of Eq.(\ref{asvv}) as 
\begin{equation}
i\int T V_{\alpha}^{a}(x) V_{\beta}^b(0) e^{ipx} d^4x |_{p^2\rightarrow\infty}\approx 
i d^{abc}\epsilon_{\alpha\beta\rho\sigma}\frac{p^{\sigma}}{P^2}
C^{NS}(A_s)~A^{c}_{\rho}(0)+ \dots
\label{ansvv}
\end{equation}
Taking now the  limit $q^2\rightarrow \infty$ in                           
Eq.(\ref{asvv})  we get  the following  
Crewther-type identity  in the SI channel  
\begin{equation} 
\label{DSI}
C_{SI}(A_s)\times C_D^{SI}(A_s)\equiv 1~~~.
\end{equation}
It  should be compared with the classical NS  Crewther identity, namely 
\begin{equation} 
\label{DNS}
C_{NS}(A_s)\times C_D^{NS}(A_s)\equiv 1~~~~.
\end{equation}
It follows   from the  $x$-space studies  
of the NS AVV three-point function \cite{Crewther:1972kn} 
(see  \cite{Adler:1973kz} as well). 
In the momentum space it was re-derived in \cite{Gabadadze:1995ei}
by considering the same three-point function of Eq.(\ref{anvv}). 
Note, that  $C_D^{SI}(A_s)$ and $C_D^{NS}(A_s)$ 
are the  coefficient functions of the massless axial-vector and vector  
Adler $D$-functions, defined by taking derivative $Q^2\frac{d}{dQ^2}$ of 
the mass-independent terms in the  correlator of SI  axial-vector currents
\begin{equation}
\label{ASI}
i\int<0|T A_{\mu}(x) A_{\nu}(0)|0>e^{iqx}d^4 x = 
\Pi^{SI}_{\mu\nu}(Q^2)=(g_{{\mu}{\nu}}q^2-q_{\mu} q_{\nu})\Pi^{SI}(Q^2)
\end{equation}
and of the correlator of NS axial-vector currents 
\begin{equation}
\label{ANS}
i\int<0|T A_{\mu}^{(a)}(x) A_{\nu}^{(b)}(0)|0>e^{iqx}d^4 x = \delta^{ab}
\Pi^{NS}_{\mu\nu}(Q^2)= \delta^{ab}(g_{{\mu}{\nu}}q^2-q_{\mu} q_{\nu})
\Pi^{NS}(Q^2)
\end{equation}
where  $Q^2=-q^2$ is the Euclidean momentum transfer. 
The  exact  chiral invariance of the 
massless perturbative expressions for  the  coefficient functions 
implies, that $C_D^{SI}(A_s)\equiv
$$C_{D}^{NS}(A_s)$. Keeping this in mind and comparing 
l.h.s. of Eq.(\ref{DSI}) and Eq.(\ref{DNS}), we get    
the following relation   
\begin{equation}    
\label{Confl}
C_{NS}(A_s)\equiv C_{SI}(A_s)|_{conformal~invariant~limit}
\end{equation}
where $A_s$ is fixed. Eq.(\ref{Confl}) is valid in the conformal-invariant 
limit of the  $SU(N_c)$ gauge 
model  and in the pqQED model in all orders of perturbative expansion   
in the fixed  expansion parameter $A=\alpha/(4\pi)$.  
In the latter case Eq.(\ref{Confl})   was proved in \cite{Kataev:2010tm}. 

\section{Conformal invariant limit of the third order perturbative series}

In the pqQED, using the detailed considerations 
of Ref.\cite{Kataev:2010tm}, it is 
possible to  demonstrate   explicitly the validity of the identity of 
Eq.(\ref{Confl})  at   level of  third order corrections. In
the process of these studies   
the following  $O(A^3)$  pqQED expressions  were used: the
order $O(A^3)$ expression for  $C_{NS}(A)$, available from   
\cite{Larin:1991tj}, and   
the defined within   dimensional regularisation \cite{'tHooft:1972fi} 
expression 
$C_{SI}(A_s)=\overline{C}_{SI}(A_s)/Z_5^{SI}(A_s)$ \cite{Larin:1994dr},
where $Z_5^{SI}$ is the finite renormalization constant of the SI 
Axial-vector current. In order to get the  
pqQED limit of all functions, contributing to $C_{SI}(A_s)$,  
in the work \cite{Kataev:2010tm} $Z_5^{SI}(A)$ was determined 
from the pqQED limit of $Z_5^{NS}(A_s)$ finite  renormalization constant, 
analytically evaluated  in  \cite{Larin:1991tj}.
Combining these inputs 
the   validity of the  identity of   Eq.(\ref{Confl}) 
at the $O(A^3)$ -approximation of pqQED was demonstrated in the 
analytical form \cite{Kataev:2010tm}. 
To fix  the  $O(A^4)$ pqQED correction to these functions  
one can  use the pqQED expression of the related 
analytical result from  \cite{Baikov:2010je}. This result coincides with 
the one, obtained in \cite{Kataev:2008sk}  from  the 
classical Crewther relation of Eq.(\ref{DNS}),    
supplemented with  the 
pqQED  $O(A^4)$ analytical approximation for  $C_{D}^{NS}(A)$, first     
presented  in   \cite{Baikov:2008cp}. The 
 $O(A^4)$ pqQED expression for $C_{NS}(A)$ reads 
\begin{equation}
\label{NSpqED}
C_{NS}(A)=1-3A+\frac{21}{2}A^2-\frac{3}{2}A^3-
\bigg(\frac{4823}{8}+96\zeta_3\bigg)A^4+O(A^5)~~.
\end{equation}  
It should coincide with the pqQED limit of still 
unknown  $O(A^4)$ coefficient of the   SI  contribution into  the 
Ellis-Jaffe sum rule.

In the case of $SU(N_c)$ gauge group with fermions 
the similar $O(A_s^3)$ expression for the NS coefficient  
functions follows from  the results of the work of Ref.\cite{Kataev:2010du}
and reads   
\begin{equation}
\label{CFTQCD}
C_{NS}(A_s)=1 -3C_F A_s+
\bigg(\frac{21}{2} C_F^2- C_FC_A\bigg) A_s^2-
\bigg[\frac{3}{2}C_F^3+65 C_F^2 C_A+\bigg(\frac{523}{12}
-216\zeta_3\bigg)C_FC_A^2 \bigg]A_s^3 
\end{equation}
It corresponds  to the conformal invariant limit of the perturbative 
result for the  $SU(N_c)$ gauge group with fermions  
and was obtained  in Ref.\cite{Kataev:2010du} 
using the the  Crewther relation of Eq.(\ref{DNS}) and 
the  the  derived  in \cite{Mikhailov:2004iq}  $\beta$-expanded  
expression for  $C_{D}^{NS}(A_s)$-function, based on the 
developed in  Ref.\cite{Mikhailov:2004iq} $\beta$-expanded 
generalization of the BLM approach, proposed in \cite{Brodsky:1982gc}.   
Here  $C_F$ and $C_A$ are the Casimir operators of the $SU(N_c)$ group. 

Taking into account the derived by us expression
of Eq.(\ref{Confl}) we conclude,  that the this expression 
should coincide with the similar approximation 
of $C_{SI}(A_s)$-contribution into Eq.(\ref{EJp}).

Note, that in the {\it conformal limit}   the ratios  of the 
corresponding perturbative  approximations  
for  the Ellis-Jaffe and  Bjorken sum rules  give us the  
following relations  
\begin{equation}
\label{EJp/Bjp}
\frac{EJ^{lp(n)}(Q^2)}{Bjp(Q^2)}=\pm \frac{1}{2}+\frac{a_8}{6~a_3}+
\frac{2\Delta \Sigma}{3 a_3} 
\end{equation}
where $a_8=3a_3-4D$,  $a3$, $a_8$ and $\Delta\Sigma$ are defined through the 
polarised parton distributions below Eqs.(\ref{EJp}) and $D$ is the hyperon 
decay constant. These  relations  coincide with the ones, 
obtained  within  massless quark-parton 
model and  can be re-written as    
$$\frac{EJ^{lp}(Q^2)}{Bjp(Q^2)}=1 +\frac{2(\Delta \Sigma-D)}{3~a_3} ~~~~;~~~~
\frac{EJ^{ln}(Q^2)}{Bjp(Q^2)}= +\frac{2}{3}\frac{(\Delta \Sigma-D)}{a_3}~~.$$ 
They lead to the standard   quark-parton model definition of the       
Bjorken sum rule through the the   Ellis-Jaffe sum rules, namely   
\begin{equation}
\label{c}
Bjp\equiv EJ^{lp}-EJ^{ln} ~~~.
\end{equation} 
Thus, our considerations  are self-consistent. 

\section{Conformal symmetry and the Bjorken 
sum rule} 

It is worth to stress, that the ``conformal invariant'' expression for 
the Bjorken sum rule  with the perturbative coefficient function defined 
in  Eq.(\ref{CFTQCD})   can be used in  phenomenological studies of  
experimental data for the Bjorken sum rule.   
This can be done with the help of  
the  principle of maximal conformality (PMC), introduced in the works of 
Ref.\cite{Brodsky:2011ig}, Ref. \cite{Brodsky:2011ta} and   
already applied in the analysis of Tevatron and LHC data in Ref. 
\cite{Brodsky:2012rj}. Within PMC principle,  one should  specify 
in the scale-dependence of the  parameter $A_s$ 
and 
substitute instead of $A_s$ its  scale-dependent 
definition $A_s(Q^{*2}_{PMC})$  into Eq.(\ref{CFTQCD}), 
leaving the  analytical 
coefficients in the related  perturbative approximation identical to those, 
obtained in the 
{\bf conformal invariant} limit of $SU(N_c)$ theory.  Note, however, that  
instead of using new scale in every new order  of perturbation theory, as was 
prescribed in Refs.\cite{Brodsky:2011ta}, \cite{Brodsky:2012rj}, 
it may be worth to use 
the unique scale $Q^{2}_{PMC}$, which should absorb all non-conformal 
invariant contributions into the expressions of the $\overline{MS}$-scheme 
coefficients of  $C_{NS}(A_s)$ coefficient 
function, defined as    
\begin{equation}
C_{NS}(A_s)=1+\sum_{l\geq 0}  c_l A_s^{l+1}(Q^2)~~.
\end{equation}  
Within the framework of the  approach of Ref. \cite{Mikhailov:2004iq} the 
$\overline{MS}$-coefficients should be expanded 
in powers of the $\beta_i$ coefficients of the 
renormalization-group $\beta$-function 
\begin{equation}
\mu^2\frac{\partial A_s}{\partial \mu^2}=-\sum_{l\geq 0} \beta_{l} A_s^{l+1}~~~.
\end{equation} 
as 
\begin{eqnarray}
\label{exp}
c_2&=&\beta_0 c_2[1]+c_2[0]~~\\ \nonumber
c_3&=&\beta_0^2c_3[2]+\beta_1c_3[0,1]+\beta_0c_3[1]+c_3[0]~~\\ \nonumber 
c_4&=&\beta_0^3c_4[3]+\beta_1\beta_0c_4[1,1]+\beta_2c_4[0,0,1]+
\beta_0^2c_4[2]+\beta_1c_4[0,1]+\beta_0c_4[1]+c_4[0]
\end{eqnarray}   
The defined in  Eq.(\ref{exp}) terms   
$c_2[1]$, $c_3[2]$, $c_3[0,1]$ and $c_3[1]$ are known  from the studies of 
Ref.\cite{Kataev:2010du}.  Notice, that  the proposed   in 
Ref.\cite{Mikhailov:2004iq},   
expansion differs a bit from the PMC realization, 
considered in Ref.\cite{Brodsky:2011ta}
by the presence in the expressions for   $c_3$ and $c_4$ coefficients   
of the additional $\beta_0c_3[1]$ known  and still explicitly  unknown 
$\beta_0^2c_4[2]$, $\beta_1c_4[0,1]$ and $\beta_0c_4[1]$  
terms, which disappear  in the {\it conformal invariant} limit.
In view of this the analogs of PMC scales, fixed from the expansion of 
Eq.(\ref{exp}),  
will differ  from the similar scales,  analogous to 
the ones, fixed in  Ref.\cite{Brodsky:2011ta} in the process 
of the analysis of the perturbative predictions for the $R(e^+e^-\rightarrow
hadrons)$.    
The similar Bjorken sum rules  studies, with more   detailed discussions  
of the   applications of 
both realizations $\beta$-expansions, which can be compared   
within  the proposed in Ref.\cite{Grunberg:1991ac} generalization of the 
original BLM approach \cite{Brodsky:1982gc},  will be considered  elsewhere 
\cite{KataevMikhailov}.

{\bf Acknowledgments.}
This is a bit more detailed  version of the  talks, presented at  the 20th International DIS-2012  
Workshop, Bonn Univ., March 26-30 ,  2012  and at 
the International Workshop ``Holography: Applications to Technicolour,
Condensed matter and Hadrons'', INR RAS, June 11-15, 2012, where 
the last section is presented in more detail.
I am grateful to the OC of DIS-2012 Workshop for the  hospitality in Bonn 
and for the  financial support.
The material of this talk was discussed during 
two weeks  stay at Ruhr Univ.  Bochum,  prior DIS Workshop. It is the 
pleasure to thank   
M.V. Polyakov and N.G. Stefanis for hospitality and  useful comments.
I am also grateful to S. J.  Brodsky, S.V. Mikhailov, O.V. Teryaev and 
D. Mueller  for the discussions of different subjects, presented in these 
talks.  The work  is done within the framework of the scientific 
program of the                                                                                                                                                                 is done within the scientific program of 
RFBR Grant N 11-01-00182 
and was supported the Grant NS-5590.2012.2.



\end{document}